\documentclass{pasj01}
\draft

\begin{document}
\Received{}%{yyyy/mm/dd}
\Accepted{}%{yyyy/mm/dd}
%\Published{yyyy/mm/dd}

\title{Rapid Dissipation of Protoplanetary Disks in Ophiuchus\thanks{Based in part on data collected at Subaru Telescope, which is operated by the National Astronomical Observatory of Japan.}}

%%% begin:list of authors
% Do NOT capitalize all letters in "textsc".
\author{Yuhei \textsc{Takagi},\altaffilmark{1}
           Yoichi \textsc{Itoh},\altaffilmark{1}
           Akira \textsc{Arai},\altaffilmark{1,2}
           Sai \textsc{Shoko},\altaffilmark{1}
           and
           Yumiko \textsc{Oasa},\altaffilmark{3}}
\altaffiltext{1}{Nishi-Harima Astronomical Observatory, Center for Astronomy, University of Hyogo, 407-2, Nishigaichi, Sayo, Sayo, Hyogo 679-5313}
\altaffiltext{2}{Koyama Astronomical Observatory, Kyoto Sangyo University, Motoyama, Kamigamo, Kita-ku, Kyoto City, 603-8555}
\altaffiltext{3}{Faculty of Education, Saitama University, 255 Shimo-Okubo, Sakura, Saitama, Saitama 388-8570}
\email{takagi@nhao.jp}
%%% end:list of authors

%%% Please use the following style in case that sorting by 
%%% affiliation is impossible. 
%
% \author{%
%   D-Firstname \textsc{D-Familyname}\altaffilmark{1}
%   E-Firstname \textsc{E-Familyname}\altaffilmark{1,2}
%   and
%   F-Firstname \textsc{F-Familyname}\altaffilmark{2}}
% \altaffiltext{1}{Address of Institute}
% \email{ddddd@xxx.xxx.xx.xx}
% \email{eeeee@xxx.xxx.xx.xx}
% \altaffiltext{2}{Address of Institute}

%% `\KeyWords{}' always has to be placed before `\maketitle'.
 %Do NOT move this preamble from here!

\maketitle

\begin{abstract}
We present the results of an age determination study for pre-main sequence stars in the Ophiuchus molecular cloud. The ages of eight pre-main sequence stars were estimated from surface gravities derived from high-resolution spectroscopy. The average age of the target stars was 0.7 Myr. By comparing the individual age and the near-infrared color excess, we found that color excess decreases gradually with a constant rate and the lifetime of the inner disk was determined to be 1.2 Myr. The estimated lifetime is nearly a half of the time compared to that of the pre-main sequence stars in the Taurus molecular cloud estimated with the same method. This result indicates that the disk evolution timescale depends on the environment of the star-forming region.
\end{abstract}

\section{Introduction}

The circumstellar disk of a pre-main sequence (PMS) star is created through the contraction of the rotating molecular gas core, where planets form. Disks dissipate due to gas and dust accretion, growth of grains, and gas dispersal driven by ultraviolet radiation. Since the formation and the evolution of planets are strongly dominated by the disk evolution timescale, an accurate estimation of this timescale is desired. 

There have been many investigations to estimate the disk evolution timescale. The early estimation was conducted by \citet{Haisch2001} who conducted the $JHKL$ photometry of six clusters. They found that 50\% of the PMS star disks dissipate in $\leq$3 Myr. This relationship was refined by \citet{Mamajek2009}. The relation between the cluster age and the disk fraction was approximated by an exponential and the e-folding time was estimated as 2.5 Myr. On the other hand, \citet{Bell2013} determined the age of the clusters using color-magnitude diagrams, and indicated a disk lifetime of 10 --- 12 Myr. Many other studies estimated the disk evolution timescale based on photometric observations (e.g., \cite{Hernandez2008}; \cite{Muzerolle2010}; \cite{Ribas2014}). From the low frequency of transitional disk objects, whose disk has an inner hole due to disk evolution, compared to PMS stars with optically thick disks, the evolution timescale of the inner disk was estimated to be of the order of $10^5$ yr \citep{Skrutskie1990}. The disk evolution timescale varies with the method used for age determination. Moreover, age determination of a PMS star based on photometric observations suffers from uncertainties in distance, extinction, and excesses arising from the accreting heated disk. 

The surface gravity, $g$, estimation of PMS stars is an alternative method to derive the age. The surface gravity of the photosphere of PMS stars decrease due to photospheric contraction over time. In our previous studies, we established surface gravity diagnostics by calculating the equivalent width ratios (EWRs) of Fe (8204.9 $\AA$) and Na (8183.3 $\AA$, 8194.8 $\AA$) absorption lines. EWRs are suitable tools to determine the surface gravity of a PMS star, as they are free from the uncertainties arising from distance, extinction, and the continuum veiling. On the other hand, we have to take into account the possibility that the EWRs of a PMS star with a high accretion rate may contain uncertainties arising from line emission filling (e.g., \cite{Petrov2011}). Previously, we observed field dwarfs and giants of which the surface gravities and effective temperatures ($T_{\mathrm{eff}}$) are well known from photometric studies and hipparcos parallaxes \citep{Takagi2010}. The relationship between the EWRs and surface gravity was established for stars with spectral types of late-K to early-M. Next, we determined the surface gravities of 10 PMS stars in the Taurus molecular cloud using the EWRs. The ages were then calculated using the evolutionary model of PMS stars \citep{Siess2000}. By comparing the estimated ages to the near infrared (NIR) color excesses, we found that the NIR color excess dissipates within several Myr \citep[hereafter Paper I]{Takagi2014}. The disk dissipation timescale was estimated from a single star forming region. To discuss the diversity of the disk dissipation timescale, we present the results of the age determination of PMS stars in the Ophiuchus molecular cloud. Similar to the Taurus molecular cloud, this is one of the closest star-forming regions, making it a suitable location to estimate the age with high-resolution spectroscopy.

\section{Observations and data reductions}

We conducted high-resolution spectroscopic observations of eight PMS stars on 2013 June 25---27 with the High Dispersion Spectrograph (HDS; \cite{Noguchi2002}) with two 2K $\times$ 4K CCD (13.5 $\micron$ square pixel), mounted on the Subaru Telescope. We selected late-K type PMS stars as targets to employ the EWR---log $g$ relation used in the previous studies (\cite{Takagi2010}, Paper I). The target stars are listed in table 1. All of the targets belong to the $\rho$ Ophiuchus ($\rho$ Oph) cloud core except V1211 Oph, which is located in the Ophiuchus North molecular cloud. Because a spectrum with high signal-to-noise ratio (S/N) is needed (S/N $\sim$ 100) to determine the surface gravity and the age, single PMS stars with $I$-band magnitude of $<$ 14 mag was selected. Since the number of late-$K$ type single PMS stars in the large survey conducted by \citet{McClure2010} is 19, our sample cover nearly one third of the known late K-type single PMS stars in this region, and almost all of the bright PMS stars suitable for the EWR method are included in present study. The optical elements of the HDS were set to an optimum value to obtain the Fe (8204.9 $\AA$) and Na (8183.3 $\AA$, 8194.8 $\AA$) lines with resolution of $\sim$60000 (0.024 $\AA$ pixel$^{-1}$). The integration time for each object was set to 600 --- 9000 s to achieve an S/N of 100 per pixel. The object which requires exposure time more than 1800 s was split into several exposures to avoid cosmic ray contamination. A0-type stars were also observed as standard stars for telluric absorption correction. Thorium-argon (Th-Ar) lamp frames were taken for wavelength calibration. 

The Image Reduction and Analysis Facility (IRAF) software package was used for the data reduction\footnote{IRAF is distributed by the National Optical Astronomy Observatory.}. Bias subtraction, flat fielding, cosmic ray rejection, and scattered light subtraction were conducted before the spectra were extracted. After wavelength calibration, telluric absorption correction, and continuum normalization, we measured the equivalent widths of the absorption lines (figure 1) using the Voigt function. The errors in equivalent widths were calculated from the uncertainties in the continuum level. The measured equivalent widths of the target stars are listed in table 1.

\begin{figure}
 \begin{center}
  \includegraphics[width=8cm]{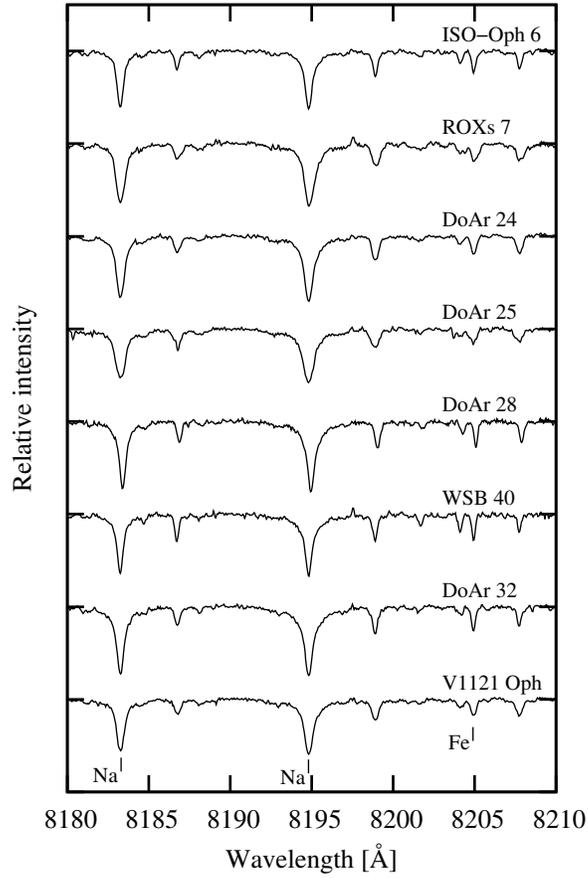} 
 \end{center}
\caption{Spectra of the targets stars. The telluric lines were removed by using the standard star.}\label{age-color}
\end{figure}

\begin{table}
\tbl{Measured equivalent widths of target stars. }{%
\begin{tabular}{lccc}
   \hline \hline
   Star 		& \multicolumn{3}{c}{EW ($\AA$)}									\\ \cline{2-4} 
   	 	& Na 8183.3$\AA$			& Na 8194.8$\AA$			& Fe 8204.9$\AA$			\\ \hline 
   ISO-Oph 6	& 0.459$^{+0.016}_{-0.017}$	& 0.572$^{+0.007}_{-0.013}$	& 0.074$^{+0.009}_{-0.004}$	\\ 
   ROXs 7	& 0.620$^{+0.028}_{-0.028}$	& 0.821$^{+0.033}_{-0.031}$	& 0.109$^{+0.009}_{-0.008}$	\\ 
   DoAr 24	& 0.604$^{+0.029}_{-0.029}$	& 0.787$^{+0.025}_{-0.022}$	& 0.082$^{+0.007}_{-0.007}$	\\ 
   DoAr 25	& 0.603$^{+0.028}_{-0.028}$	& 0.796$^{+0.025}_{-0.036}$	& 0.085$^{+0.009}_{-0.008}$	\\ 
   DoAr 28	& 0.596$^{+0.021}_{-0.013}$	& 0.779$^{+0.021}_{-0.024}$	& 0.079$^{+0.009}_{-0.004}$	\\ 
   WSB 40	& 0.492$^{+0.020}_{-0.012}$	& 0.669$^{+0.028}_{-0.016}$	& 0.080$^{+0.010}_{-0.008}$	\\ 
   DoAr 32	& 0.682$^{+0.025}_{-0.023}$	& 0.870$^{+0.031}_{-0.033}$	& 0.096$^{+0.015}_{-0.015}$	\\ 
   V1121Oph	& 0.494$^{+0.015}_{-0.016}$	& 0.672$^{+0.017}_{-0.016}$	& 0.083$^{+0.009}_{-0.008}$	\\ 
   \hline
  \end{tabular}}
\end{table}

\section{Results}
The surface gravity determinations are conducted by employing the EWR---log $g$ relationship described in Paper I. The relationship between the surface gravity and the Fe (8204.9 $\AA$) / Na (8183.3 $\AA$) and Fe (8204.9 $\AA$) / Na (8194.8 $\AA$) were estimated by observing 21 field giants and four dwarfs \citep{Takagi2010}. All of these field stars have $T_{\mathrm{eff}}$ of $\sim$4200 K, comparable to late-K type PMS stars. The surface gravity of each star was calculated as follows:
\begin{center}
\begin{equation}
\log g = \log \frac{M}{M_{\odot}} + 4\log \frac{T}{T_{\odot}} - \log \frac{L}{L_{\odot}} + \log g_{\odot},
\end{equation}
\end{center}
where $M$, $T$, and $L$ are the mass, effective temperature, and luminosity, respectively. We used the parallax in \citet{vanLeeuwen2007} and the $V$-band magnitude from Tycho-2 catalog \citep{Hog2000} for the luminosity calculations. The Tycho $V$ magnitude was translated to Johnson $V$ magnitude before the surface gravity calculation \citep{ESA1997}. The mass of each giant star was determined by comparing the location in the HR diagram and the theoretical evolutionary tracks \citep{Lejeune2001}. The masses of dwarfs were estimated by using the spectral type - mass relation described in \citet{Drilling2000}.  As mentioned in Paper I, the EWR---log $g$ relationship is expressed as:
\begin{equation}
  \log g = \frac{\log EWR - \log a}{b}, 
\end{equation}
where $g$ is measured in units of cm $\mathrm{s}^{-1}$. For Fe (8204.9 $\AA$) / Na (8183.3 $\AA$), the values of a and b are 1.522$\pm$0.040 and -0.289$\pm$0.004, respectively. For Fe (8204.9 $\AA$) / Na (8194.8 $\AA$), the values are 1.141$\pm$0.024 and -0.294$\pm$0.004, respectively. All of the coefficients and their errors are estimated by employing the weighted Deming regression\footnote{These coefficients are updated from Paper I due to the magnitude correction of Tycho-2 catalog and adopting the Deming method.}. 
The error of the surface gravity was calculated with the following expression: 
\begin{equation}
  \delta \log g = \sqrt{\left(-\frac{1}{ab\ln 10}\delta a\right)^2+\left(\frac{\log a - \log EWR}{b^2}\delta b\right)^2+\left(\frac{1}{bEWR\ln 10}\delta EWR\right)^2}. 
\end{equation} 
Derived surface gravities are listed in table 2, which range from 3.2 to 3.6 dex. These gravities are estimated from EWRs, which show the state of the photosphere and are free from uncertainties in distance, extinction and continuum veiling. In addition, the weighted average of  log $g$ is 3.42, and the weighted average of log $g$ of the four stars with log $g <$ 3.42 is 3.29$\pm$0.06, and that of the rest four star is 3.58$\pm$0.06. These points indicate the dispersion of log $g$ and the age spread in a single star forming region.

The age of each target was estimated by comparing the derived surface gravity to the PMS evolutionary model described by \citet{Siess2000}. For the calculation, the $T_{\mathrm{eff}}$ of each target was determined by comparing the spectral types quoted from \citet{Wilking2005}, \citet{Andrews2007}, and \citet{McClure2010} to the spectral type - $T_{\mathrm{eff}}$ relation described by \citet{Kenyon1995}. The results are listed in table 2. The mass of a target star is estimated by interpolating the evolutionary tracks. The values listed in table 2 do not include the errors in the $T_{\mathrm{eff}}$ of the targets. The estimated ages may include an additional error of $\sim$20\%, by assuming a spectral type uncertainty of $\pm$1 subclass. This is relatively small compared to the error arising from the S/N of the spectrum and the fitting error of EWR---log $g$. The mass of a PMS star will vary depending on the applied evolutionary model \citep{Hillenbrand2004}. However, the relation between the age and the surface gravity is comparable in PMS stars around 1 $M_{\odot}$. Age estimations based on surface gravity is less dependent of the evolutionary models of PMS stars. Figure 2 shows the H-R diagram with the plots of the target stars. 

\begin{table*}
 %\begin{center} 
  \tbl{Estimated properties of the target stars. }{%
   \begin{tabular}{lccccccc}
   \hline \hline
   Star 		& Type\footnotemark[$*$]	& SpT	& $T_{\mathrm{eff}}$ 	& log $g$			& $M_{\rm star}$	& Age	& log $L_{\rm star}/L_{\odot}$	\\ 
		&	&		    & (K)					&			& ($M_{\odot}$) 				& (Myr)	& \\ \hline
ISO-Oph6	& C	& K4.5	 & 4470	 & 3.30$^{+0.07}_{-0.14}$	& 1.54$^{+0.15}_{-0.07}$	& 0.43$^{+0.07}_{-0.12}$	& 0.94$^{+0.17}_{-0.09}$	\\ 
ROXs7	& W	& K5.5	 & 4280	 & 3.22$^{+0.09}_{-0.11}$	& 1.09$^{+0.06}_{-0.04}$	& 0.40$^{+0.08}_{-0.08}$	& 0.79$^{+0.14}_{-0.11}$	\\ 
DoAr24	& C	& K5	 & 4350	 & 3.58$^{+0.11}_{-0.11}$	& 1.12$^{+0.03}_{-0.00}$	& 0.91$^{+0.33}_{-0.24}$	& 0.47$^{+0.13}_{-0.11}$	\\ 
DoAr25	& C	& K5	 & 4350	 & 3.54$^{+0.12}_{-0.13}$	& 1.13$^{+0.02}_{-0.01}$	& 0.82$^{+0.32}_{-0.24}$	& 0.51$^{+0.14}_{-0.12}$	\\ 
DoAr28	& TD	& K5	 & 4350	 & 3.63$^{+0.08}_{-0.13}$	& 1.11$^{+0.03}_{-0.00}$	& 1.05$^{+0.26}_{-0.32}$	& 0.42$^{+0.14}_{-0.08}$	\\ 
WSB40	& C	& K5.5	 & 4280	 & 3.35$^{+0.11}_{-0.15}$	& 1.04$^{+0.05}_{-0.02}$	& 0.53$^{+0.15}_{-0.15}$	& 0.64$^{+0.17}_{-0.12}$	\\ 
DoAr32	& C	& K6	 & 4205	 & 3.52$^{+0.17}_{-0.17}$	& 0.90$^{+0.02}_{-0.00}$	& 0.82$^{+0.48}_{-0.27}$	& 0.37$^{+0.19}_{-0.16}$	\\ 
V1121Oph	& C	& K5	 & 4350	 & 3.31$^{+0.11}_{-0.13}$	& 1.20$^{+0.08}_{-0.05}$	& 0.47$^{+0.13}_{-0.11}$	& 0.77$^{+0.16}_{-0.13}$	\\ 
    \hline
  \end{tabular}}
 %\end{center}
\begin{tabnote}
\hangindent6pt\noindent
\hbox to6pt{\footnotemark[$*$]\hss}\unskip%
 C, W, and TD correspond to classical T Tauri star (CTTS), weak-lined T Tauri star (WTTS) and transitional disk objects quoted from
 \citet{Wilking2005}, \citet{Andrews2007}, and \citet{McClure2010}. DoAr 28 was classified as TD based on the SED in \citet{McClure2010}.
\end{tabnote}
\end{table*}

The weighted average age of the targets is 0.7 Myr. Due to the brightness limit of observations, the target stars selected in the present study are concentrated at the edge of clouds where the extinction is small. The median age of the PMS stars located in this surface region of the $\rho$ Oph core is estimated as 2.1 Myr \citep{Wilking2005}. The age estimated in this work is less than a half of this result. It is more comparable to the age estimations of the young stars in the core part of the $\rho$ Oph cloud, which is estimated as 0.3 Myr (e.g., \cite{Greene1995}).   

\begin{figure}
 \begin{center}
  \includegraphics[width=8cm]{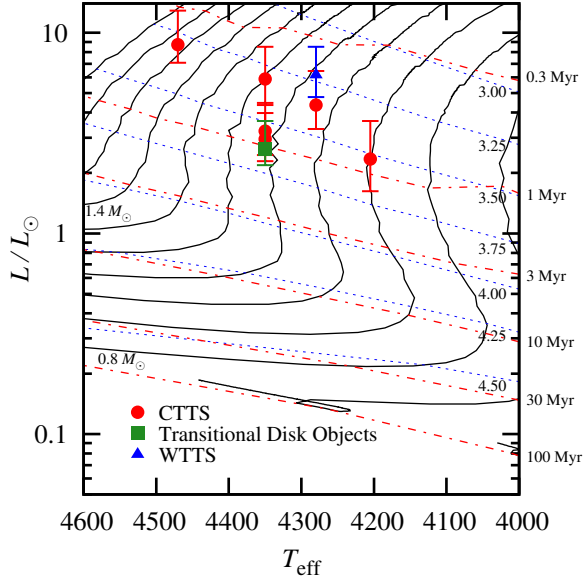} 
 \end{center}
\caption{HR diagram of the observed stars. The luminosity of each target were estimated from the comparison of the log $g$ and the evolutionary model of PMS stars \citep{Siess2000}. The solid lines are evolutionary tracks (Siess et al. 2000) for masses of 0.4 to 1.7 $M_{\odot}$. The red dot-dashed lines and the blue dotted lines indicate isochrones and isogravity lines.}\label{age-color}
\end{figure}

To discuss the disk dissipation timescale in the Ophiuchus cloud, we estimated the NIR color excess ($J-K$) arising from the heated disk. The NIR excess of a PMS star arises from the inner part of the disk where it is optically thick. A large NIR excess indicates that the hot disk lies near the central star. As the inner edge of the disk spreads outward, the NIR excess will decrease due to the lack of hot disk. The NIR $J$ and $K$ magnitudes of the targets were taken from the results of the 2MASS survey \citep{Skrutskie2006}. The intrinsic color arising from the photosphere was subtracted by quoting the value from \citet{Kenyon1995}. The reddening caused by interstellar material was removed by adopting the reddening raw (\cite{Cohen1981}, \cite{Meyer1997}) using the value of the extinction ($A_V$) estimated in previous studies. Extinctions were preferentially quoted from the study which conducted extinction corrected spectral energy distribution (SED) fittings mainly based on near-infrared observations (e.g., \cite{McClure2010}), in order to compare with the result of Paper I which also used $A_V$ calculated from SED estimations. $A_V$ was calculated using the extinction law of \citet{Mathis1990} when the extinction value in $V$-band was not presented. The adopted NIR color excesses are listed in table 3.

\begin{table}
 \tbl{The NIR color excesses of the target stars.}{%
   \begin{tabular}{lccc}
   \hline \hline
   Star 		&  $J-K$		& $A_V$	& Ref.\footnotemark[$*$]	\\ 
		&  (mag)		& 	& 			\\ \hline
   ISO-Oph 6	& 0.55$\pm$0.03	& 2.2	& 1			\\ 
   ROXs 7	& 0.10$\pm$0.04	& 6.0	& 2			\\ 
   DoAr 24	& 0.26$\pm$0.04	& 3.5	& 1			\\ 
   DoAr 25	& 0.23$\pm$0.03	& 3.4	& 1			\\ 
   DoAr 28	& 0.15$\pm$0.03	& 2.3	& 1			\\ 
   WSB 40	& 0.49$\pm$0.03	& 4.3	& 1			\\ 
   DoAr 32	& 0.18$\pm$0.03	& 5.9	& 3			\\ 
   V1121Oph	& 0.43$\pm$0.05	& 1.1	& 4			\\ 
   \hline
  \end{tabular}}
 \begin{tabnote}
  \hangindent6pt\noindent
  \hbox to6pt{\footnotemark[$*$]\hss}\unskip%
  $A_V$ are quoted from; (1) \cite{McClure2010}; (2) \cite{Wilking2005}; (3) \cite{Cieza2010}; (4) \cite{Sartori2003}.
 \end{tabnote}
\end{table}

\begin{figure}
 \begin{center}
  \includegraphics[width=8cm]{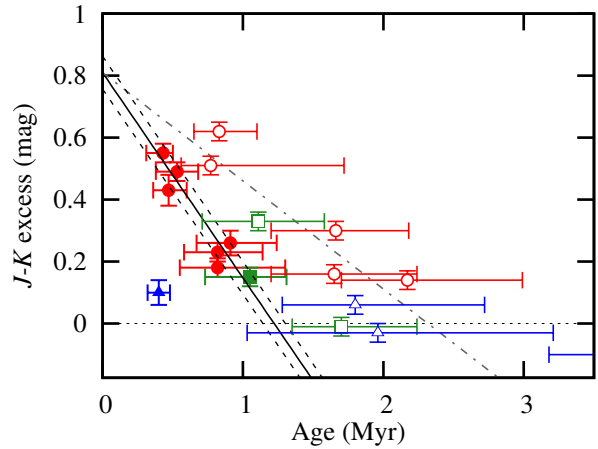} 
 \end{center}
\caption{The age - $J-K$ excess relationship of the observed stars. Filled plots are the same as figure 2. Open plots of red circles, green square, and blue triangles represent Class II objects, transitional disk objects, and Class III objects of Taurus molecular clouds, respectively, with recalculations of the results in Paper I (see the text). The solid line indicates the regression line derived from the plots without ROXs 7. The dashed lines are drawn based on the 1 $\sigma$ uncertainty of the regression line at $J-K$ excess = 0. The gray dotted-dashed line shows the relationship for PMS stars in the Taurus molecular cloud calculated with the stars with $J-K > 0$.}\label{age-color}
\end{figure}

We derived the disk dissipation timescale in the Ophiuchus molecular cloud by comparing the estimated ages with the NIR color excesses (figure 3). The $J-K$ excess decays at a constant rate. ROXs 7, which is the only WTTS included among our targets, shows a small excess despite of its youth. The presence of this object may indicate that the PMS stars in this regions have different disk lifetime, or a rapid disk dissipation occurred in a specific PMS star. On the other hand, atomic lines of Na, Fe and V in 8100---8200 $\AA$ show an asymmetric profile, which imply the presence of cool spots or a companion star. This profile was not found in spectra of other targets. The estimated age of ROXs 7 may include an additional error from this reason. 

We estimated the typical disk dissipation timescale of this region without ROXs 7. The disk dissipation timescale estimated from the $J-K$ color excess is 1.2$\pm$0.1 Myr. Errors were estimated from the standard deviations of the weighted regression line coefficients. Because the targets of the present work is limited to PMS stars with small extinctions, the disk dissipation timescale of the stars in the core part of the cloud could be different. 
Since PMS stars in the Ophiuchus clouds are young ($\sim$1 Myr), there is a lack of stars with $J-K$ excess $\approx$ 0. Therefore there is a possibility that the tail part of the decline of age-color excess relationship will extend  to older ages. Observations of faint disk-less PMS stars with next generation telescopes or a near-infrared observations of the PMS stars deeply embedded in the molecular cloud can reveal more accurate disk dissipation timescale and also the specificity of ROXs 7. 

It should be noted that the age and NIR color excess estimation depends on the employed relationships between spectral type, $T_{\mathrm{eff}}$, and intrinsic colors. We also calculated the age and NIR color by using the relation of \citet{Pecaut2013} and compared with the result shown in table 2. The age of each PMS stars becomes 17\% older on average, although they are still within the range of error. In addition, the $J-K$ excess of each star will be 0.1 mag smaller. Nevertheless, the estimated disk dissipation lifetime using this relation is 1.2$\pm$0.1 Myr, which is comparable to the result mentioned above.

\section{Discussions}

The near-infrared excesses indicate that the disk exists in the inner region of the disk. The decrease of the NIR color excess implies the outspreading inner hole. \citet{Ribas2014} estimated the disk evolution timescale by comparing the age of the 22 young clusters and the fraction of the PMS stars with near-infrared and/or mid-infrared excesses. The timescales estimated from the shorter wavelength and the mid-infrared were 2---3 Myr and 4---6Myr, respectively. They suggest that the dust evolution in the inner disk is more rapid than that in the outer part. The estimated timescale in the present study indicates more rapid disk clearing in the Ophiuchus molecular cloud. 

The dissipation timescale of PMS star disks in the Taurus molecular cloud estimated with the EWR method and the NIR magnitudes from 2MASS was 2.3$\pm$0.2 Myr\footnote{This timescale is recalculated by using the observational results of Paper I with the revised EWR-log $g$ relationship described in section 3. The ratio of Sc and Na in the near-infrared $K$-band was also used for the age estimations of PMS stars in the Taurus cloud. The values of a and b in equation (2) for Sc (22057.8$\AA$) / Na (22062.4 $\AA$) are 2.464$\pm$0.051 and -0.289$\pm$0.003, respectively. For Sc (22057.8$\AA$) / Na (22089.7 $\AA$), the values of a and b are 2.987$\pm$0.065 and -0.276$\pm$0.004, respectively.} (figure 3). The estimated timescale in the Ophiuchus molecular cloud in the present study is nearly a half of time compared to the timescale in Taurus. The difference in disk dissipation timescale between the Ophiuchus and Taurus molecular clouds indicates that the disk evolution timescale will varies between star-forming regions. 

The gradual disk dissipation with a constant rate is consistent with disk dispersal triggered by the disk wind driven by magnetorotational instability (MRI) -driven turbulence \citep{Suzuki2010}. The vertical magnetic fields in the disk are amplified by the MRI and the gas in the surface layer of the disk will flow outward. The constant decline of the NIR excess can be explained with this wind by assuming that dust grains are well coupled with the gas. The strength of this disk wind is determined by the strength of the magnetic flux. The masses of PMS stars in the Taurus and Ophiuchus molecular clouds used for the disk dissipation timescale estimations in Paper I and the present study are $\sim$1 $M_{\odot}$. Thus, the magnetic flux density of their parent cores is considered to be nearly equal. The difference between the disk dissipation timescales may imply that the fractions of magnetic field diffusion during star and disk formation are different between the Taurus and Ophiuchus molecular clouds.

Another possible reason for the difference in disk dissipation timescales is the presence of high-mass stars. While the Taurus molecular cloud is in a static environment, the Ophiuchus molecular cloud is placed close to the Sco-Cen OB association. The radiation pressure and stellar wind from massive stars increase the density and the magnetic field of the molecular cloud \citep{Kusune2015}. This may suggest that the molecular cloud cores in Ophiuchus collapsed with a stronger magnetic flux compared to those in Taurus. \citet{Wilking2005} suggested that the surface layer of the Ophiuchus molecular cloud was stripped away by the ultraviolet radiation of massive stars. This radiation may have the potential to dissipate the disks, as in the circumstellar disks in Orion (e.g., \cite{O'Dell1994}). 

The difference in metallicity may also be responsible for the unique disk dissipation timescale for each star-forming region. Both observational (e.g., \cite{Yasui2010}) and theoretical studies (e.g., \cite{Ercolano2010}) investigated the changes in evolution timescale caused by the differences in metallicity, and concluded that low metallicity can result in rapid disk dissipation. On the other hand, PMS stars in the Taurus molecular cloud and the Ophiuchus molecular cloud have metallicity nearly equivalent to the Sun \citep{Padgett1996}. As this study was performed using a hypothetical value of the surface gravity, further investigations of the metallicities based on accurate surface gravity and veiling estimations may reveal the cause of the timescale difference between the Taurus and Ophiuchus molecular clouds. 

In addition, one may consider other factors for determining the NIR excess. When the dust sublimation radius is smaller than the magnetospheric radius, the inner edge of the disk may be controlled by the position of co-rotation and hence the rotation period of the star. On the other hand, if the sublimation radius lies beyond the magnetosphere, the position of the inner edge can be managed by the radiation field. Moreover, the shape and the inclination of the disk can have an impact on the NIR color (e.g., \cite{Thi2011}).
Revealing these parameters will contribute to the better understanding of the disk evolution and its timescale.

\bigskip

This research was based on data collected at the Subaru Telescope, which is operated by the National Astronomical Observatory of Japan. This work was supported by JSPS KAKENHI Grant Numbers 26103708, 26870507, and 25870124.

%%%
% See the manual for the detail.
%%%

\end{document}